\documentclass[twocolumn,showpacs,preprintnumbers,amsmath,amssymb]{revtex4}

\font\mybb=msbm10 at 10pt
\def\bb#1{\hbox{\mybb#1}}

\def\bE {\bb{E}}

\def\tr{{\rm tr}}
\def\Tr{{\rm Tr}}
\newcommand{\be}{\begin{equation}}
\newcommand{\ee}{\end{equation}}
\newcommand{\bea}{\begin{eqnarray}}
\newcommand{\eea}{\end{eqnarray}}
\newcommand{\ba}{\begin{array}}
\newcommand{\ea}{\end{array}}

\def\bbox{{\,\lower0.9pt\vbox{\hrule \hbox{\vrule height 0.2 cm
\hskip 0.2 cm \vrule height 0.2 cm}\hrule}\,}}
\newcommand{\dsl}{\pa \kern-0.5em /}

\usepackage{graphicx}
\usepackage{dcolumn}
\usepackage{bm}

\begin{document}

\preprint{DAMTP-2006-42, UG-06-06, hep-th/0607193}

\title{Open M5-branes }

\author{Eric A. Bergshoeff}
\affiliation{%
Centre for Theoretical Physics, University of Groningen,\\
Nijenborgh 4, 9747 AG Groningen, The Netherlands.
}%

\author{Gary W. Gibbons and Paul K. Townsend}
\affiliation{
Department of Applied Mathematics and
Theoretical Physics,\\
Centre for Mathematical Sciences, University of Cambridge,\\
Wilberforce Road, Cambridge, CB3 0WA, U.K.
}%


\begin{abstract}

We show how, in heterotic M-theory, an M5-brane in the 11-dimensional
bulk may end on an  ``M9-brane'' boundary, the M5-brane boundary 
being a Yang monopole 4-brane. This possibility suggests various novel 
5-brane configurations of heterotic M-theory, in particular a static 
M5-brane suspended between the two M9-brane boundaries, for which we 
find the asymptotic heterotic supergravity solution.

\end{abstract}

\maketitle

At the semi-classical level, M-theory is a theory of 11-dimensional supergravity interacting with 
its 1/2 supersymmetric branes, such as M2-branes and M5-branes, but also `M9-branes'  \cite{Horava:1996ma},  which are actually boundaries that support a 10-dimensional supersymmetric $E_8$ gauge theory. Central to many applications of M-theory  is an understanding of  how some M-branes may have boundaries on others. This is in  part due to the implications for
superstring theories when viewed as M-theory compactifications. For example, the possibility of D-branes in type II string theory \cite{Polchinski:1995mt} can be deduced from the fact that M2-branes may have boundaries on M5-branes \cite{Strominger:1995ac,Townsend:1995af}. Another example is the interpretation of the $E_8\times E_8$ heterotic string as an M2-brane suspended between two M9-branes  \cite{Horava:1996ma}. 

There is a further possibility for open branes that was recently emphasized by Polchinski in the context of heterotic string theories \cite{Polchinski:2005bg}. He has shown that  the Green-Schwarz anomaly cancellation term implies that the $SO(32)$ heterotic string can have an endpoint on a `monopole', defined as a point such that  the integral of the 8-form $\Tr F^4$ over an 8-sphere enclosing it is non-zero, where $F$ is the Yang-Mills (YM) field-strength 2-form. This possibility is `non-standard'  because the `monopole' has infinite energy, which means that the string endpoint is not free to move in the 9-dimensional space;  the string can ``end but not break''.  Analogous ``non-standard''  open D-branes have recently been discussed \cite{Bergman:2006aa}. 

Here we show that M5-branes may have boundaries on M9-branes. The boundary 4-brane has infinite tension, so its centre of mass is not free to move, although it may fluctuate. For a planar boundary 
4-brane one can ignore the  four directions of the brane;   from this perspective the M5-brane boundary is a Yang monopole  \cite{Yang:1977qv}, which can be defined generally as a singular point of $SU(2)$ YM fields in 5-space (ignoring time) for which the integral of $\Tr F^2$ over  any 4-sphere enclosing it is  
non-zero (with $SU(2) \subset E_8$ here).  This possibility allows several novel 5-brane configurations,  such as the suspension of an M5-brane between two M9-branes, for which we find the asymptotic form of the corresponding heterotic supergravity solution.  

We begin with the bosonic truncation of the effective heterotic  supergravity theory,
without  the quantum anomaly cancelation term. This  involves the 10-dimensional supergravity fields (metric $g$, dilaton $\phi$, and 2-form potential $B$ with 3-form field strength $H$) and the 
Yang-Mills gauge fields.  If all fermion fields are omitted, 
the Lagrangian density is
\be\label{density}
{\cal L} =  \sqrt{-g}\, e^{-2\phi}\left[R +4|\partial\phi|^2 -
\frac{1}{2}|H|^2
- \kappa\,  \Tr |F|^2\right], 
\ee
where the trace is taken in the adjoint representation, and $\kappa$
is  a constant proportional to the inverse string tension. The 3-form
field strength for $B$ is $H= dB + \kappa\, \omega$, 
where $\omega$ is the  Chern-Simons 3-form satisfying $d\omega= \Tr F^2$, 
as a consequence of which $H$ satisfies the `anomalous' Bianchi identity
\be\label{Bianchi}
dH = \kappa\, \Tr F^2\, . 
\ee

A heterotic string is an electric source for $B$ with a string charge  
$Q_1 \propto \int\! e^{-2\phi} \star H$, where the 7-form $\star H$ is the Hodge dual of $H$, and the integral is over any 7-sphere threaded by the string. Because the equations of motion imply that $e^{-2\phi}\star H$ is a closed form, this 7-sphere can be deformed  arbitrarily, without 
changing the value of the integral, as long as no singularities of $H$ are crossed. The presence of a heterotic string implies a singularity of $H$ at the string core, but if the string were to have an endpoint we could contract the 7-sphere to a point without crossing this singularity, and thereby deduce that $Q_1=0$. Thus free heterotic string endpoints are forbidden, classically, although new possibilities can arise as a consequence of anomaly cancellation terms  \cite{Polchinski:2005bg}. 

In addition to strings, heterotic string theories also have 5-branes. 
A planar 5-brane carries a 5-brane 
charge 
\be
Q_5 = \frac{1}{32\pi^2 \kappa} \int_{S^3_\infty} H\, , 
\ee
where the integral is over the 3-sphere at transverse spatial infinity. There are two possible 
contributions to this integral, corresponding to two types of 5-brane. For the `solitonic' 5-brane \cite{Strominger:1990et}, which we will call the H5-brane, the 3-sphere at infinity can be contracted to a point on a transverse 4-space $\Sigma$ on which all fields are non-singular.  In this case, the Bianchi identity (\ref{Bianchi}) yields
\be\label{q5}
Q_5= \frac{1}{32\pi^2}  \int _\Sigma \Tr F^2\,  . 
\ee
We shall focus on the $E_8\times E_8$ theory and assume that the 
YM fields of the H5-brane take values in the Lie algebra of the $SU(2)$ factor of an  
$E_7\times SU(2)$ subgroup of one of the $E_8$ factors of $E_8\times E_8$. 
In this case $Q_5$ is the $SU(2)$ instanton number. The worldvolume
dynamics of a one-instanton H5-brane is governed by an action
involving 30 worldvolume hypermultiplets \cite{Callan:1991dj}. 

The other type of 5-brane \cite{Duff:1990wv} can be found by shrinking the H5 instanton
to zero size. In this process the region in which the YM fields are
non-zero shrinks to a point, but this point simultaneously recedes to
infinite affine distance as an infinite `throat'  forms in which the
spacetime becomes the product of a 3-sphere of fixed radius and a
7-dimensional `linear-dilaton' vacuum
\cite{Callan:1991dj,Gibbons:1993sv}. In this limit, the YM fields 
become gauge-equivalent to the zero-field configuration, 
leaving a purely gravitational `black'  5-brane. However, a `large'
gauge transformation is needed to gauge away the YM fields and this
requires a transformation of $B$ that transforms $H$ into a closed but
non-exact 3-form with a non-zero integral over the 3-sphere, such that
the charge $Q_5$ remains the same. The linearly increasing dilaton
implies a strong coupling limit, indicating that this `black' 5-brane
is best understood from the M-theory perspective. In fact, it is just
the M5-brane interpreted as a solution of the minimal  10-dimensional
supergravity \cite{Duff:1994fg}, and the above process in 
which the H5-brane is converted 
into a  `black'  5-brane can be interpreted as one in which a 5-brane 
is pulled from the 10-dimensional boundary into the 11-dimensional 
bulk \cite{Ganor:1996mu}. 

Now imagine that the above process, converting an H5-brane into an M5-brane, 
is carried out not in time but in some space direction, so that the instanton core 
of a planar H5-brane goes to zero size everywhere on a 4-plane. Since the 
4 planar directions play no role we may periodically identify in these directions 
and consider the same process for a string in an effective 6-dimensional 
heterotic supergravity theory. Omitting the supergravity fields we have an 
instanton string in which the core shrinks to zero size at some point, resulting in a 
Yang monopole endpoint. To see this,  consider the topological 4-sphere that results 
from the union of a 4-ball in $\Sigma$ with a hemi-4-spherical `cap'  of the same
radius $R$, in the limit of large $R$. As $F=0$ on the cap, the integral of $\Tr F^2$
over this 4-sphere reduces to the integral over $\Sigma$, which is $32\pi^2 Q_5$, so 
the instanton string ends on a Yang monopole of strength $Q_5$.  
In the context of heterotic M-theory, the instanton string endpoint becomes the 4-dimensional interface between an M5-brane in the bulk and an H5-brane 
in the M9-brane boundary. In effect, the M5-brane boundary is a Yang-monopole  4-brane, but one with only $SO(4)$ symmetry rather than the $SO(5)$ 
symmetry of Yang's original solution of the YM equations on $\bE^5$ \cite{Yang:1977qv}. 
This is expected since no non-zero value of the 3-form $H$ could be compatible with $SO(5)$ 
symmetry. 

Let us now suppose that we have an M5-brane suspended across 
an interval bounded by two M9-branes, 
with a Yang monopole 4-brane boundary on each of the two M9-branes. From a 10-dimensional perspective, we now have {\it two} Yang monopoles, one in an $SU(2)$ subgroup of one $E_8$ and another in an $SU(2)$ subgroup of the other $E_8$. Interpreting 
the Yang monopole as a semi-infinite H5-brane would allow two possible $SO(4)$-invariant
configurations, as shown in Figs. 1 and 2.  Fig. 1 illustrates a
configuration that has a 10-dimensional interpretation as a 5-brane
ending on a `double' Yang monopole of monopole `number' $(1,-1)$, the
first entry being the instanton number for the Yang monopole on one
boundary and the second entry being the instanton number for the Yang
monopole on the other boundary. The sign is determined by the
contribution to the 5-brane charge $Q_5$, which must vanish for this
configuration. We thus have an open 5-brane of the heterotic string
theory, which will be unstable against the formation of `holes' with
$(1,-1)$ Yang-monopole boundaries. Fig. 2 illustrates a
configuration that has a 10-dimensional interpretation as a `kink' on
an H5-brane; on either side we have an H5-brane with $Q_5=1$, but on
one side this arises from instanton `number' $(1,0)$ and on the other
side from instanton `number' $(0,1)$. As the YM flux is `incoming'  for one H5-brane 
and `outgoing'  for the other one, the kink 4-brane is again a Yang
monopole of monopole `number' $(1,-1)$.

\begin{figure}
\begin{minipage}[t]{0.5\linewidth}
\centering
\includegraphics[scale=.4]{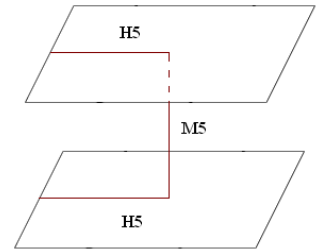}
\caption{An open 5-brane} \label{fig:Right}
\end{minipage}
\end{figure}

\begin{figure}
\begin{minipage}[t]{0.5\linewidth}
\centering
\includegraphics[scale=.4]{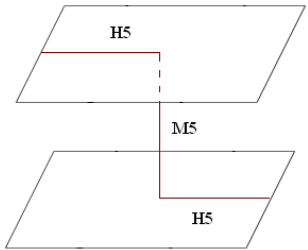}
\caption{A 5-brane `interface'} \label{fig:Left}
\end{minipage}
\end{figure}

We noted above that no non-zero $H$ is compatible with $SO(5)$
symmetry, but now that we have
two Yang monopoles it is possible to arrange for $H$ to vanish. 
From the Bianchi identity  (\ref{Bianchi}) we see that $H=0$ requires
\be
\Tr_1 F^2 + \Tr_2 F^2 =0
\ee
where the subscripts indicate a trace over the adjoint of the Lie 
algebra of either the `first' or the `second' $E_8$ factor of the 
$E_8\times E_8$ gauge group. This constraint is satisfied by 
coincident $SO(5)$ invariant Yang monopoles with  monopole `number'  
$(1,-1)$. Such a solution represents a static M5-brane suspended 
between two M9-branes, as shown in Fig. 3.  The 4-brane boundary on 
each M9-brane is now a {\it spherically-symmetric} Yang monopole. 
Individually both are sources for $H$ but their contributions cancel. 

These considerations lead us to seek a static planar 4-brane solution of heterotic supergravity 
that is a source for a $(1,-1)$ Yang monopole. We can find this solution by a lift to 10 dimensions of a
corresponding solution of the effective 6-dimensional theory obtained
by toroidal compactification. Using the ansatz 
\be
ds^2_{10} = e^{\sigma/2} ds^2_6 + e^{-\sigma/2} ds^2(T^4)\, ,
\qquad \phi = \sigma\, , 
\ee
one finds a consistent truncation to 6-dimensional gravity, in Einstein frame, coupled to a  6-dimensional dilaton field $\sigma$, a 3-form field strength, and the   $E_8\times E_8$ YM fields.  A further truncation of the 3-form field strength is   inconsistent, in general, but for the special solution we seek it can be consistently set to zero. We may also consistently truncate each of the $E_8$ multiplets of YM fields  to an $SU(2)$ triplet. The resulting 6-dimensional Lagrangian density is
\be
{\cal L}_6 = \sqrt{-\det g} \left[ R - (\partial\sigma)^2 - 
4\kappa e^{-\sigma}\, \tr |F|^2\right]  
\ee
where the gauge group is now $SU(2)\times SU(2)$ and the trace is
taken in the fundamental $({\bf 2}, {\bf 1}) \oplus ({\bf 1},{\bf 2})$
representation; this explains the additional factor of $4$. 
We now seek spherical symmetric solutions of this model that
generalize the self-gravitating Yang monopole solutions of
 \cite{Gibbons:2006wd} to include the dilaton. 
Spherical symmetry implies that the metric takes the form
\be
ds^2_6 = - e^{2\lambda(r)}\Delta(r) dt^2 + dr^2/\Delta(r) + r^2d\Omega_4^2
\ee
in terms of two functions $\lambda$ and $\Delta$, where $d\Omega_4^2$
is the $SO(5)$ invariant metric on the unit $4$-sphere. It is convenient to set
\be
\Delta(r) = 1- 2\mu(r)/r^3
\ee 
for `mass function' $\mu(r)$.  In the absence of the $\sigma$ field this model has a self-gravitating Yang monopole solution for each $SU(2)$ subgroup of the $SU(2)\times SU(2)$ gauge group. Because the YM field strength 2-form for this solution has components only on the 4-sphere, it continues to solve the YM equations as modified by the dilaton. One thus finds that
\be
\tr |F^2| = 2\times 3/r^4
\ee
where we have used the result of \cite{Gibbons:2006wd}  for the
$SU(2)$ self-gravitating Yang monopole and the factor of $2$ arises
from the necessity to consider a $(1,-1)$ Yang monopole. The Einstein
and dilaton equations then reduce to 
the equations 
\bea
\Delta r^4\sigma'{}' + 12\kappa e^{-\sigma}&& \nonumber \\
+ \left(4r^3 - 2\mu - 6\kappa e^{-\sigma}r\right) \sigma' &=& 0\, ,\\
\mu'   -  3\kappa e^{-\sigma} - \frac{1}{8}(\sigma')^2 r^4 \Delta &=& 0\, ,
\label{eqtwo}
\eea
which are a pair of coupled ODEs for  $\mu(r)$ and $\sigma(r)$, and the one further equation 
\be
4\lambda' =  r (\sigma')^2 \, , 
\ee
which can be solved for $\lambda(r)$, given $\sigma(r)$, up to an irrelevant integration constant which we choose such that $\lambda(\infty)=0$.  It is useful to note that these three equations imply the $\sigma$ equation of motion
\be\label{eofm}
\left(e^\lambda  \Delta r^4 \sigma'\right)'  = 
- 12\kappa e^{-\sigma}e^\lambda\, .
\ee

We have not found an explicit solution to these  equations but one can find an asymptotic solution of the form
\bea
\sigma &=& \sigma_0 + \frac{A}{ r^2} + \frac{\Sigma}{r^3}  + \frac{A^2}{ 2 r^4} + \dots \\
\mu &=& \frac{Ar}{2} + \mu_0 - \frac{A\Sigma}{ 2r^2} - \frac{4A^3 + 9\Sigma^2}{ 24 r^3} + \dots
\eea
where $\sigma_0$, $\mu_0$ and $\Sigma$ are arbitrary constants, and 
\be
A= 6\kappa e^{-\sigma_0}\, . 
\ee
The asymptotic solution for $\lambda$ is then found to be
\be
\lambda= -\frac{A^2}{ r^4} - \frac{3A\Sigma}{ 5r^5} + \dots
\ee
The integration constant $\sigma_0$ determines the string coupling constant $g_s= e^{\sigma_0}$ (since the zero mode of the 10-dimensional dilaton $\phi$ equals $\sigma_0$). The integration constant $\mu_0$ can be identified as a 
`Schwarzschild' mass (which may be negative because the total mass is infinite). The integration constant $\Sigma$ can be identified as a scalar charge, which could also be negative since it is manifest from the expansion for $\sigma$ that there
is also a linearly divergent contribution to this charge proportional to $\kappa$. Note that it is consistent to set $\mu_0=\Sigma=0$ and in this case the asymptotic expansion has the property that $\mu(r)$ is an odd function of $r$ and $\sigma(r)$ an even function. 

\begin{figure}
\begin{minipage}[t]{0.5\linewidth}
\centering
\includegraphics[scale=.4]{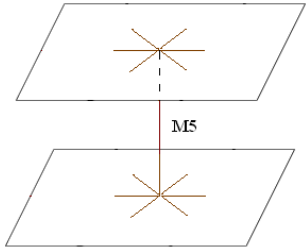}
\caption{An open M5-brane} \label{fig:OpenM5}
\end{minipage}
\end{figure}

Having fixed the asymptotic behaviour by the choice of integration 
constants, the equations determine the behaviour  in the interior.  
There cannot be a regular origin at $r=0$ because regularity of
$\Delta$ would imply $\mu'(0)=0$ and then (\ref{eqtwo}) requires
$\sigma(0)=\infty$. There could be an event horizon, at $r=r_H>0$ such
that $\Delta(r_H)=0$. In the absence of a Yang monopole (for which the
equations are as above but with $\kappa=0$)
this is possible only if $\sigma$ is constant, because then (\ref{eofm})
implies that $e^\lambda  \Delta r^4 \sigma'$ is a constant, which 
must vanish if all fields are regular as $\Delta\to 0$. Solutions of 
the Einstein-YM equations with a horizon exist in the presence of a 
Yang monopole \cite{Gibbons:2006wd}, and an analysis of the 
conditions for a regular horizon in the present case shows that the integration 
constant $\Sigma$ is determined in terms of $A$, so that $\sigma_0$ and $\mu_0$
are the free parameters. This leads us to conjecture that there exist solutions 
of the Einstein-YM-dilaton equations considered here that are
regular on and outside an event horizon. A detailed analysis would be needed
to determine whether such a solution is stable, but  we expect stability 
against splitting into spatially-separated $(1,0)$ and  $(0,-1)$ Yang-monopoles
because the interaction between them is entirely (super)gravitational.
We also expect marginal stability against collapse to a configuration of the type 
shown in Figs. 1 and 2 because  the energy  of a spherically 
symmetric  Yang monopole within a ball of  radius $r$ is $E(r)=Tr$, where $T$ is  precisely the instanton-string tension \cite{Gibbons:2006wd}, which is the H5-brane tension in the 
current context. 

Finally, we wish to point out that a membrane suspended between the
two M9-branes could end on the M5-brane suspended between the two
M9-branes. From the 10-dimensional perspective, this would appear to
be an $E_8\times E_8$ heterotic string with an endpoint on a $(1,-1)$
Yang monopole. However, it is unclear to us what happens to the chiral
modes of the string at this endpoint. Conceivably they leave the
string in a manner that is analogous to that described by Polchinski
for an endpoint of an $SO(32)$ heterotic string \cite{Polchinski:2005bg}, but in that case the
mechanism involved quantum anomaly considerations that are not
obviously relevant here. We suspect that a proper understanding of 
open M5-branes and any M2-branes that end on them will involve
quantum M-theory considerations. 

In conclusion, we have provided a concrete realization of Yang monopoles in M-theory
that may open up for investigation a new class of stable non-supersymmetric brane 
configurations in string theory. Previous discoveries of this nature have led to important insights 
into quantum field theories, and one may hope for similar insights from open M5-branes.

\begin{acknowledgments}
EAB is supported by the European Commission FP6 program MRTN-CT-2004-005104, in which he is associated to Utrecht University. PKT thanks the  EPSRC for financial suport, and the university of Groningen for hospitallity.

\end{acknowledgments}


\begin{thebibliography}{99}

 
\bibitem{Horava:1996ma}
P.~Ho{\v r}ava and E.~Witten,
Nucl.\ Phys.\ B {\bf 475}, 94 (1996)
[arXiv:hep-th/9603142];
  Nucl.\ Phys.\ B {\bf 460} (1996) 506
  [arXiv:hep-th/9510209].


\bibitem{Polchinski:1995mt}
  J.~Polchinski,
  Phys.\ Rev.\ Lett.\  {\bf 75} (1995) 4724
  [arXiv:hep-th/9510017].
  
\bibitem{Strominger:1995ac}
  A.~Strominger,
  Phys.\ Lett.\ B {\bf 383} (1996) 44
  [arXiv:hep-th/9512059].
  
\bibitem{Townsend:1995af}
  P.~K.~Townsend,
  Phys.\ Lett.\ B {\bf 373} (1996) 68
  [arXiv:hep-th/9512062].
  
\bibitem{Polchinski:2005bg}
  J.~Polchinski,
  JHEP {\bf 0609} (2006) 082
  [arXiv:hep-th/0510033].
    
\bibitem{Bergman:2006aa}
  O.~Bergman and G.~Lifschytz,
  arXiv:hep-th/0606216.
  
\bibitem{Yang:1977qv}
C.~N.~Yang,
J.\ Math.\ Phys.\  {\bf 19}, 320 (1978).
 

\bibitem{Strominger:1990et}
A.~Strominger,
Nucl.\ Phys.\ B {\bf 343}, 167 (1990)
[Erratum-ibid.\ B {\bf 353}, 565 (1991)].

\bibitem{Callan:1991dj}
  C.~G.~Callan, J.~A.~Harvey and A.~Strominger,
  Nucl.\ Phys.\ B {\bf 359} (1991) 611;
Nucl.\ Phys.\ B {\bf 367}, 60 (1991).


\bibitem{Duff:1990wv}
  M.~J.~Duff and J.~X.~Lu,
  Nucl.\ Phys.\ B {\bf 354} (1991) 141.
  
   
\bibitem{Gibbons:1993sv}
G.~W.~Gibbons and P.~K.~Townsend,
Phys.\ Rev.\ Lett.\  {\bf 71}, 3754 (1993)
[arXiv:hep-th/9307049].

\bibitem{Duff:1994fg}
M.~J.~Duff, G.~W.~Gibbons and P.~K.~Townsend,
Phys.\ Lett.\ B {\bf 332}, 321 (1994)
[arXiv:hep-th/9405124].

\bibitem{Ganor:1996mu}
O.~J.~Ganor and A.~Hanany,
Nucl.\ Phys.\ B {\bf 474}, 122 (1996)
[arXiv:hep-th/9602120].

\bibitem{Gibbons:2006wd}
  G.~W.~Gibbons and P.~K.~Townsend,
  Class.\ Quant.\ Grav.\  {\bf 23} (2006) 4873
  [arXiv:hep-th/0604024].

 

  
 
 
  
   


  



\end{thebibliography}
\end{document}